\documentclass{PoS}

\title{Search for neutral and charged MSSM Higgs bosons in ATLAS}

\ShortTitle{Search for neutral and charged MSSM Higgs bosons in ATLAS}

\author{\speaker{Martin FLECHL}
         \thanks{on behalf of the ATLAS Collaboration.}\\
        Universit\"at Freiburg\\
        E-mail: \email{martin.flechl@cern.ch}}

\abstract{
The results of Higgs boson searches in the context of the Minimal Supersymmetric extension of the 
Standard Model (MSSM) in proton-proton collisions with the ATLAS detector 
based on collected data corresponding to up to 36 pb$^{-1}$ are presented. Searches in the 
channels $H^+ \to c\bar{s}$, $H^+ \to \tau\nu$, and $H \to \tau\tau$ are discussed. 
All observations agree with the expectation of the Standard Model (SM)-only hypothesis and 
thus exclusion limits are derived.
}

\FullConference{The 2011 Europhysics Conference on High Energy Physics-HEP 2011,\\
July 21-27, 2011\\
Grenoble, Rh\^one-Alpes France}

\begin{document}
\section{Introduction}
\noindent
The Higgs sector of the MSSM consists of two complex scalar fields leading to the 
existence of five physical Higgs bosons, three of them neutral ($h$, $H$, $A$), and a charged pair ($H^\pm$). 
Searches for these particles have been performed in the $H^+ \to c\bar{s}$,
$H^+ \to \tau\nu$ and $h/H/A \to \tau\tau$ channels with the ATLAS detector~\cite{atlas}. 
%and the data are found to be compatible with the Standard Model.
The recommendations of the LHC Higgs cross section working group~\cite{xsec} have been 
used for signal rates.% and their uncertainties.
\section{Charged Higgs boson decays to $c\bar{s}$}
\begin{figure}[!h!tpb]
  \begin{center}
  \includegraphics[width=0.49\textwidth]{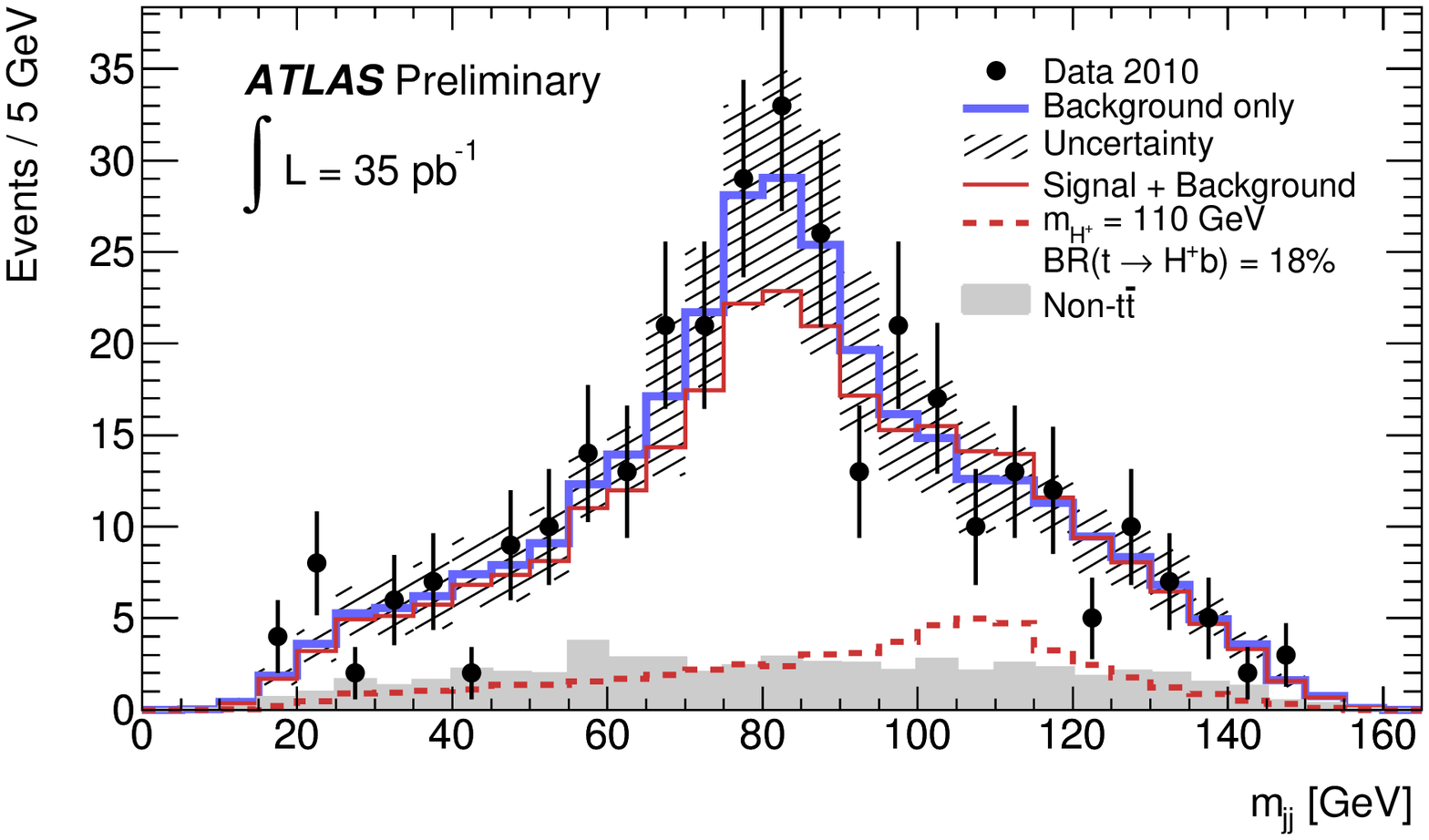}
  \includegraphics[width=0.49\textwidth]{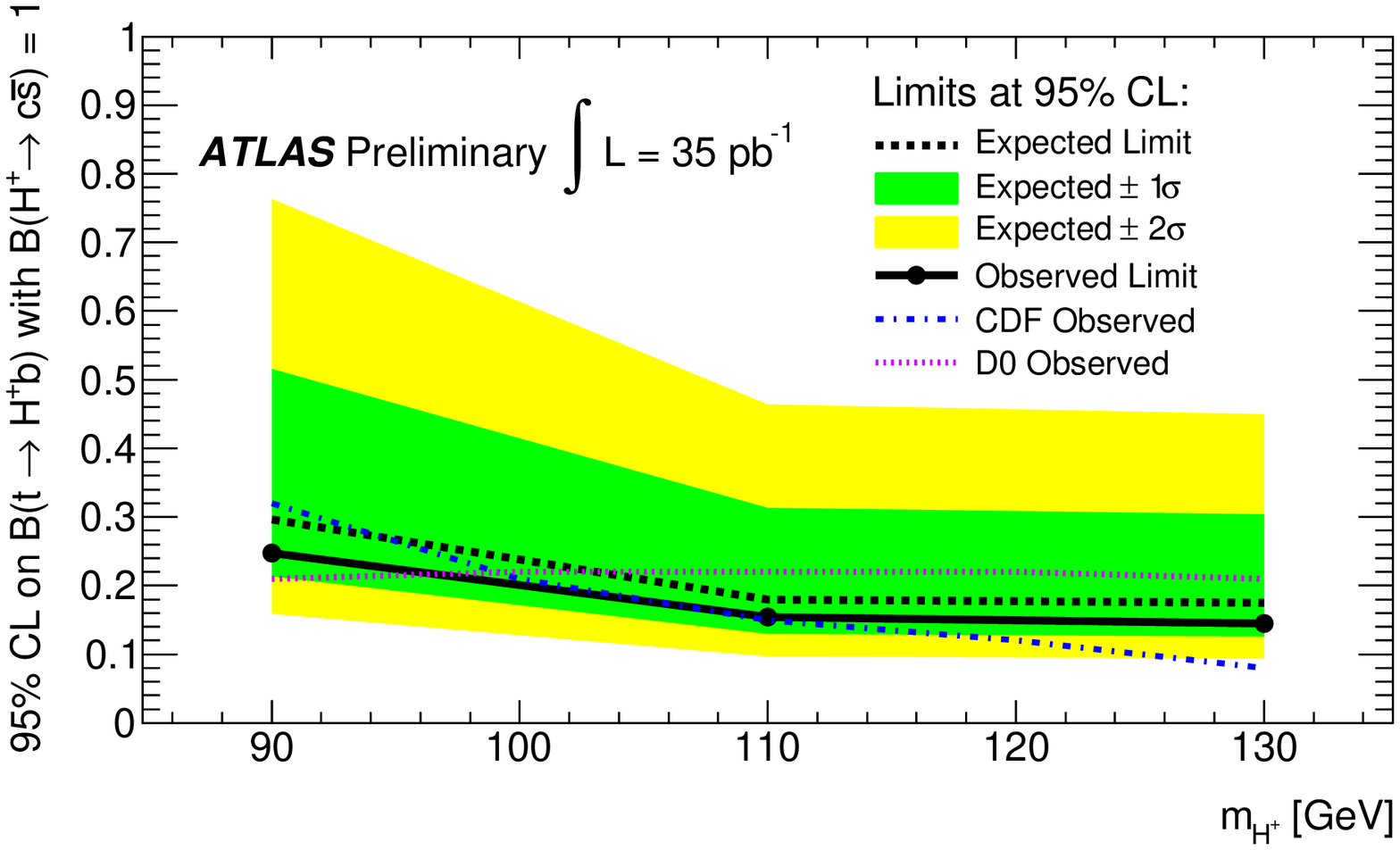}
  \end{center}
\vspace{-0.7cm}
  \caption{
Left: The dijet mass distribution after lepton+jets event selection for data is compared to 
the SM-only and the signal expectation with BR($t \to bH^+$)=0.18 and BR($H^+ \to c\bar{s}$)=1. 
Right: 95\% C.L. limit on BR($t \to bH^+$) using the CLs~\cite{cls} method.}
\label{fig:cslimit}
\end{figure}

\noindent
The channel $H^+ \to c\bar{s}$ is important for light charged Higgs bosons and $\tan\beta < 3$. The 
ATLAS search focuses on production in $t\bar{t}$ events in the decay mode $b\ell\nu bH^+$~\cite{cs}.
The signal topology is the same as for SM $t\bar{t}$ events in the lepton+jets mode but the 
invariant mass of the light-flavour jet pair corresponds to the $H^+$ instead of the $W$ mass. The 
analysis strategy is to select lepton+jets events by requiring an isolated lepton, significant missing 
transverse energy ($E_T^{\mathrm{miss}}$), at least 4 jets of which at least one is $b$-tagged, and 
a transverse mass $m_T$(lepton, $E_T^{\mathrm{miss}}$) compatible with a $W$ boson, 
and to look for a second peak in the resulting dijet mass distribution shown in Fig.~\ref{fig:cslimit}.
A kinematic $\chi^2$ fitter is applied to the whole event, and the multi-jet background is estimated 
from data. As the event yield and shape agrees with the SM expectation, limits at the 95\% C.L. 
are set (see Fig.~\ref{fig:cslimit}). A branching ratio BR($t \to bH^+$) of 0.25-0.14 is excluded 
in the mass range 90-130 GeV, assuming BR($H^+ \to c\bar{s}$)=1.
\section{Charged Higgs boson decays to $\tau\nu$}
\noindent
In the MSSM, the decay $H^+ \to \tau\nu$ has a branching ratio close to 1 for $\tan \beta > 3$ and $m_{H^+}<m_t$. 
Four channels have been studied in the context $t\bar{t} \to bW bH^+$, $H^+ \to \tau\nu$: 
$W \to qq$, hadronic $\tau$ decays (``$\tau$+jets''),
$W \to \ell\nu$, hadronic $\tau$ decays (``$\tau$+lepton'')~\cite{tauhad}; 
$W \to qq$, leptonic $\tau$ decays (``lepton+jets''),
$W \to \ell\nu$, leptonic $\tau$ decays (``dilepton'')~\cite{taulep}. 
For all searches, the main background are SM $t\bar{t}$ decays. Other backgrounds considered 
are $W/Z$+jets, single-top, and multi-jets. 

\begin{figure}[!h!tpb]
  \begin{center}
  \includegraphics[width=0.36\textwidth]{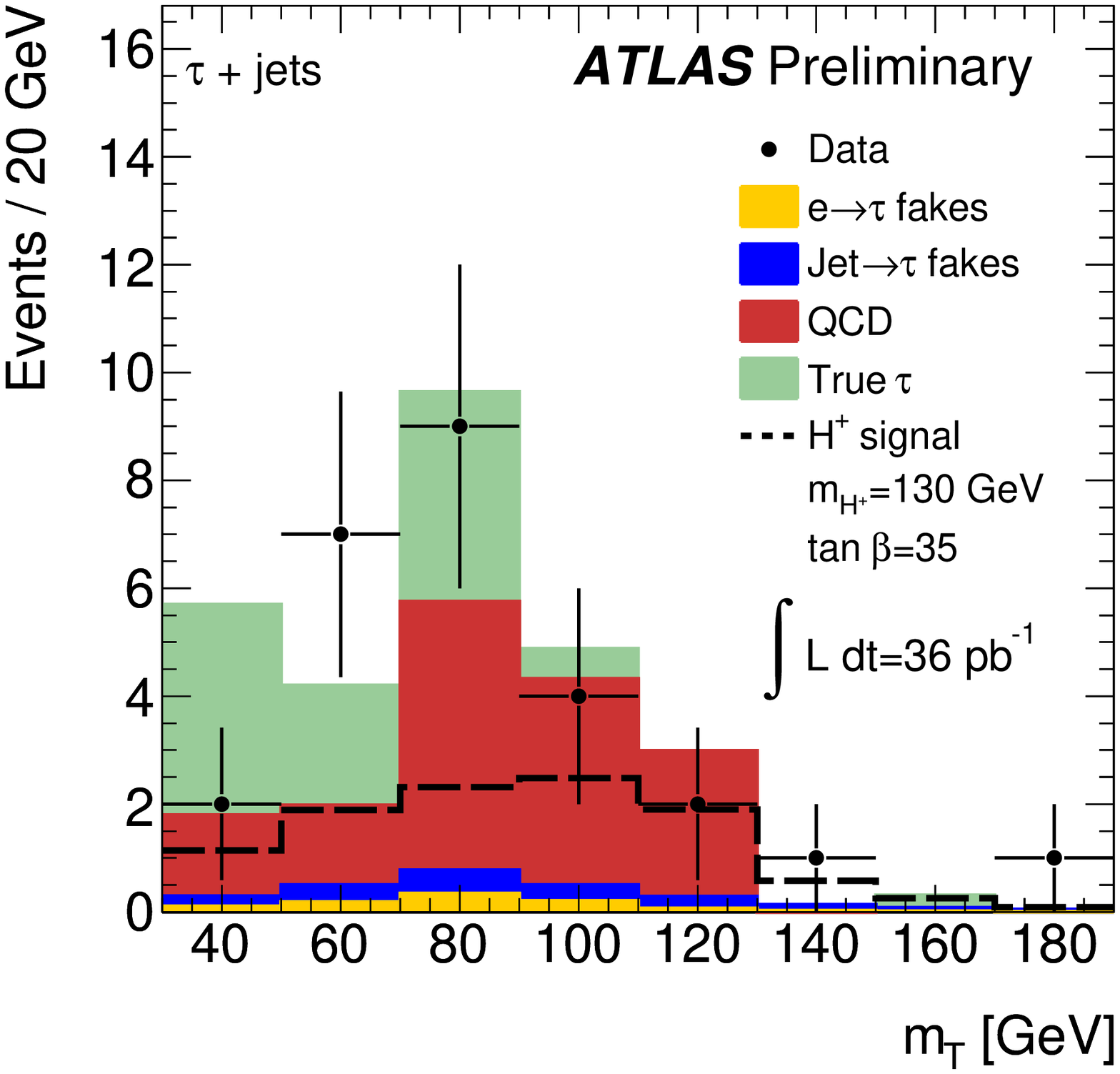}
  \includegraphics[width=0.36\textwidth]{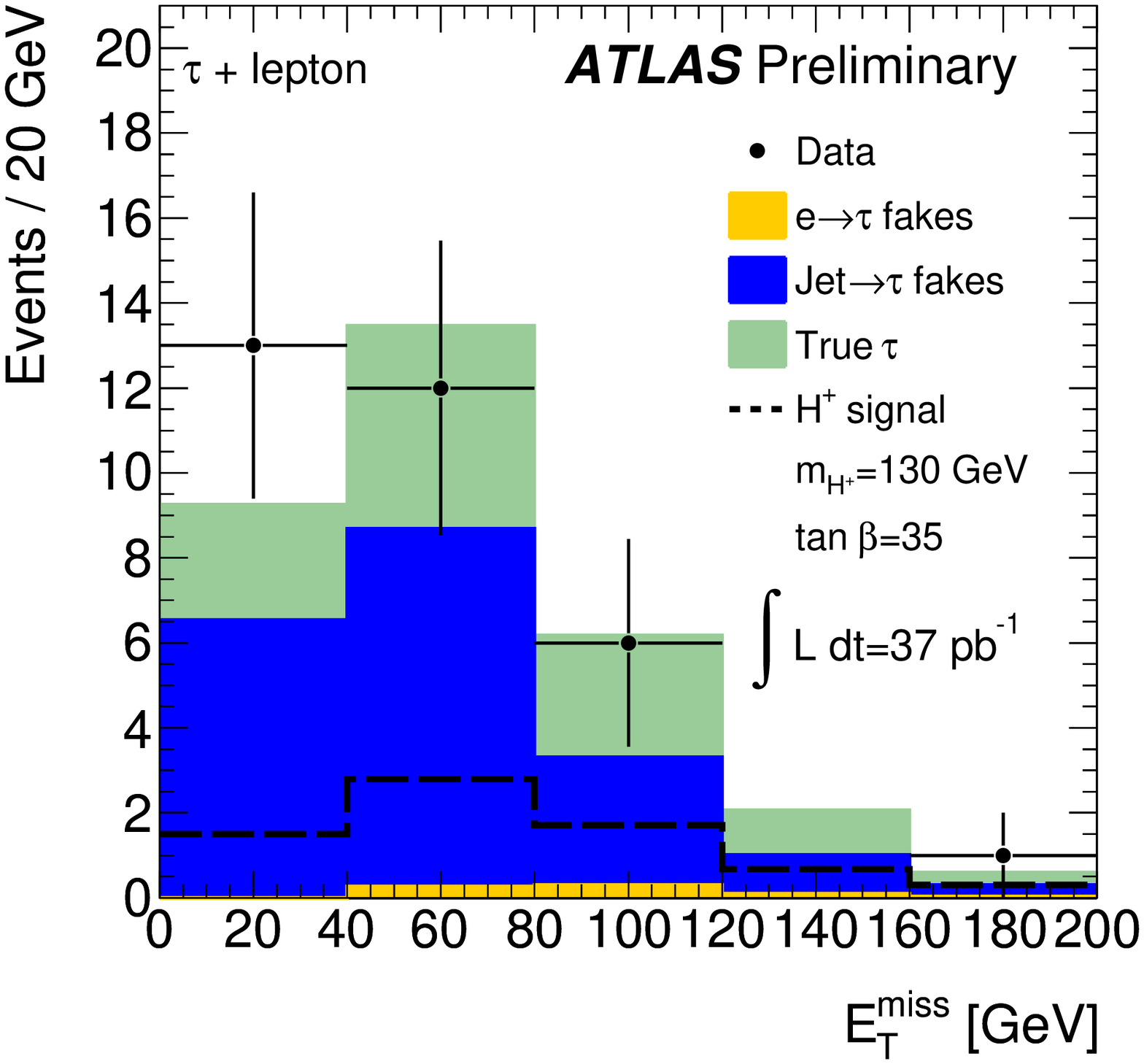}
  \end{center}
\vspace{-0.7cm}
  \caption{
Left: The $m_T$ distribution after event selection for the $\tau$+jets channel. 
Right: The $E_T^{\mathrm{miss}}$ distribution after event selection for the $\tau$+lepton channel. 
The distribution of the $H^+$ signal is given for a reference point in parameter space 
corresponding to BR$(t \rightarrow b H^+) \approx 6\%$. Backgrounds in which electron and jets are 
misidentified as $\tau$ jets, $\tau$ jets are identified correctly, and QCD multi-jet are 
shown separately.
}
\label{fig:tauhad}
\end{figure}

\begin{figure}[!h!tpb]
  \begin{center}
  \includegraphics[width=0.37\textwidth]{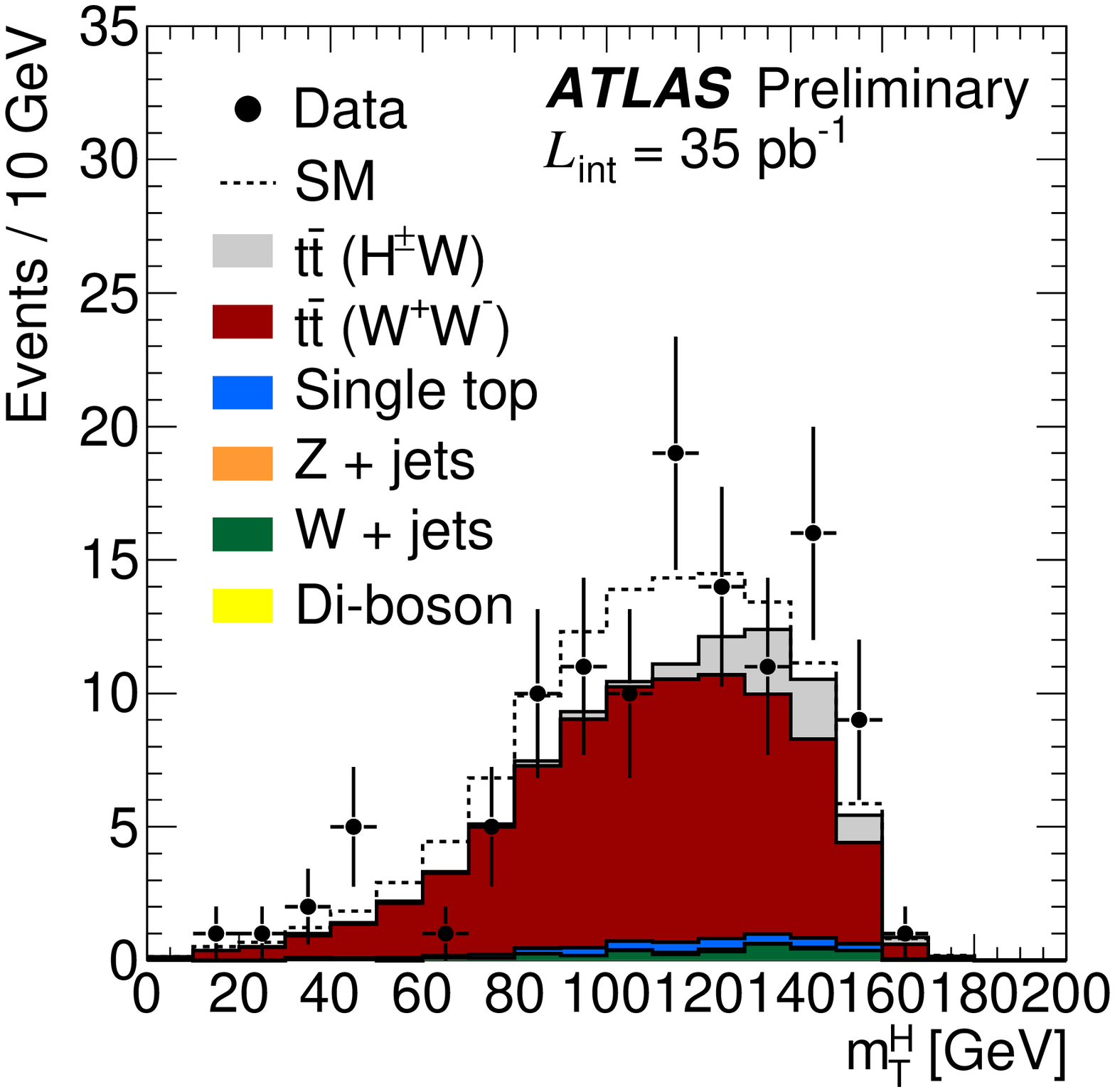}
  \includegraphics[width=0.37\textwidth]{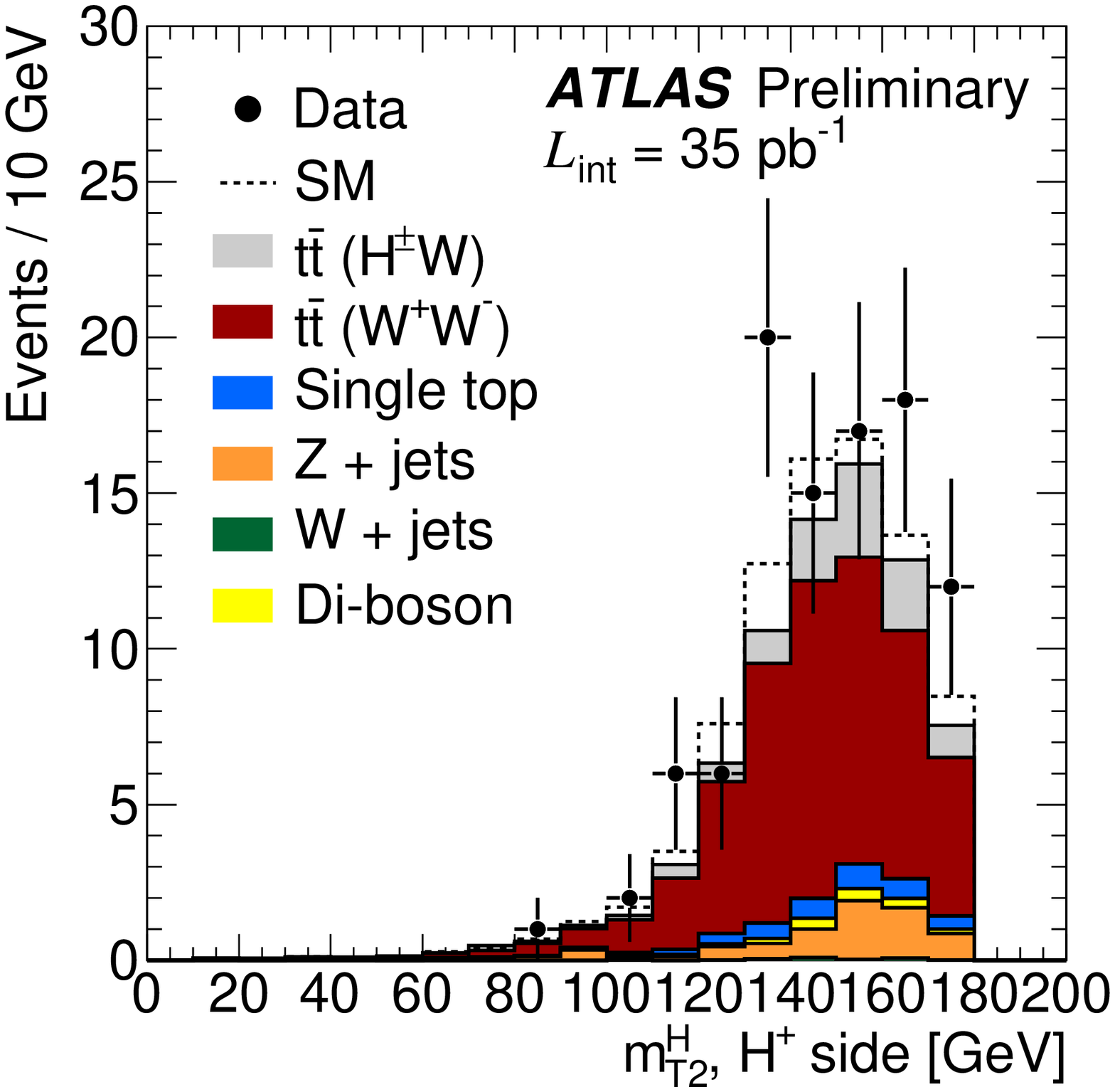}
  \end{center}
\vspace{-0.7cm}
  \caption{
Left: The $m_T^H$ distribution after event selection for the lepton+jets channel. 
Right: The $m_{T2}^H$ distribution after event selection for the dilepton channel. 
The dashed line shows the SM expectation, and the stacked histograms the 
signal+background expectation for BR($t \to bH^+$)=0.15.
}
\label{fig:taulep}
\end{figure}

The $\tau$+jets event selection requires one $\tau$ jet, no electrons or muons in the event, 
significant $E_T^{\mathrm{miss}}$, at least four jets of which at least one is $b$-tagged, 
and a top quark candidate reconstructed; and the $\tau$+lepton one $\tau$ jet, one electron or 
muon with opposite charge with respect to the $\tau$ jet, at least two jets of which at least one is $b$-tagged, 
and the transverse energy sum to be larger than 200 GeV. The final discriminants 
for the analyses involving $\tau$ jets are shown in Fig.~\ref{fig:tauhad}.
For the lepton+jets searches, one isolated electron or muon is required, at least four jets 
of which two are $b$-tagged, significant $E_T^{\mathrm{miss}}$, and a top quark candidate. 
The dilepton analysis requires two oppositely-charged isolated leptons, at least two jets. 
For same-flavor events, $|m_{ll}-m_{Z}|>10$ GeV and significant $E_T^{\mathrm{miss}}$ is required; 
otherwise a transverse energy sum larger than 150 GeV. The final generalized 
transverse mass distributions~\cite{mass} are shown in Fig.~\ref{fig:taulep}. 
Within statistics, a good agreement with the SM expectation is observed for all channels.
%\cite{newtauhad}
%
\section{Neutral MSSM Higgs boson decays to $\tau\tau$}
\noindent
The most promising channel for the observation of neutral Higgs bosons in the context of the MSSM 
is the decay to two $\tau$ leptons. The gluon fusion and $b$-associated production modes are considered, 
and ATLAS searches focus on the channels where one $\tau$ lepton decays hadronically, and the other 
leptonically (``lepton-hadron'') or where both $\tau$ leptons decay leptonically (``lepton-lepton'')~\cite{tautau}. 
The dominant background is $Z/\gamma^* \to \tau\tau$; $W$+jets, $t\bar{t}$, diboson, and multi-jet 
backgrounds have been considered as well.

\begin{figure}[!h!tpb]
  \begin{center}
  \includegraphics[width=0.32\textwidth]{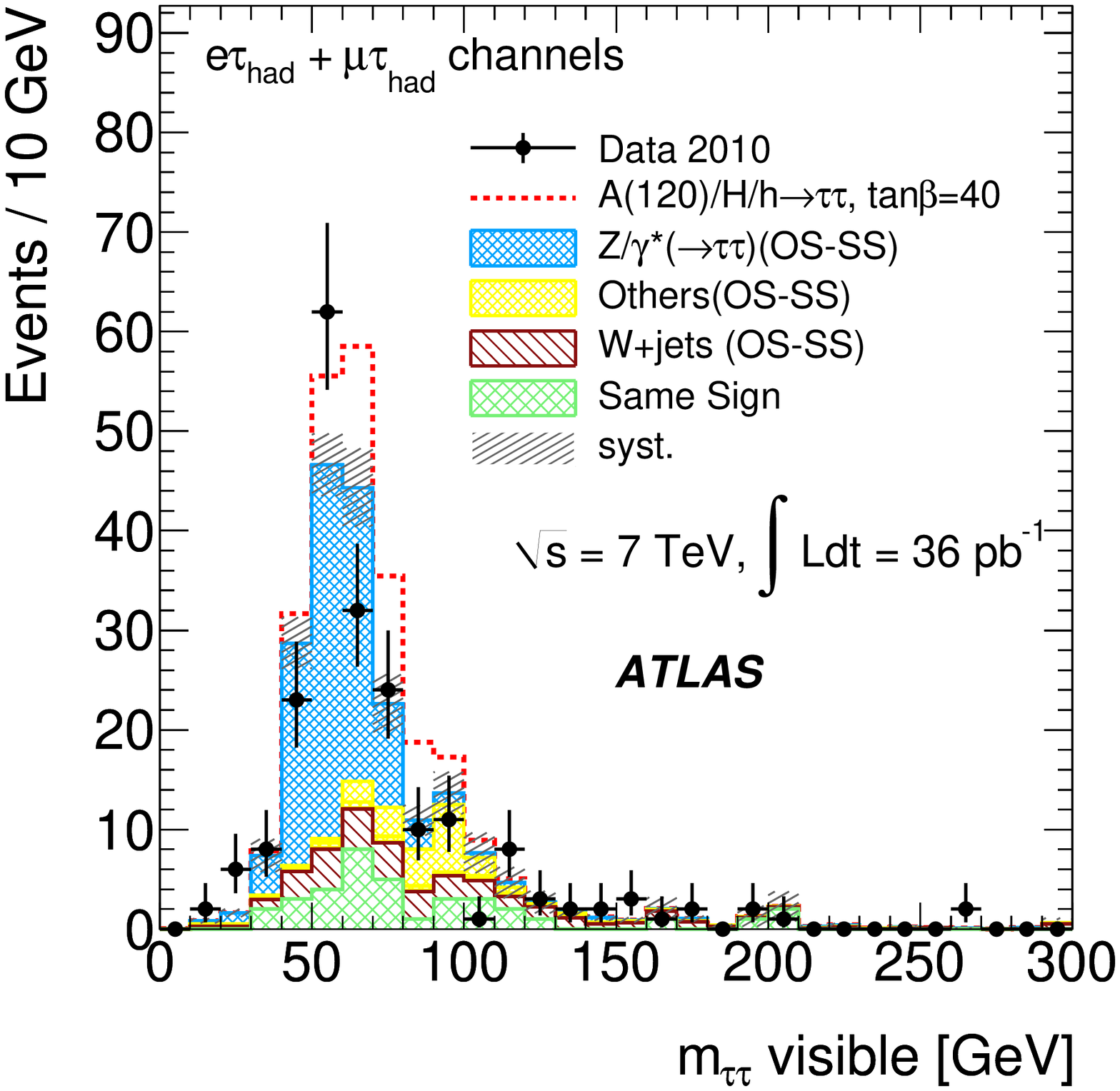}
  \includegraphics[width=0.32\textwidth]{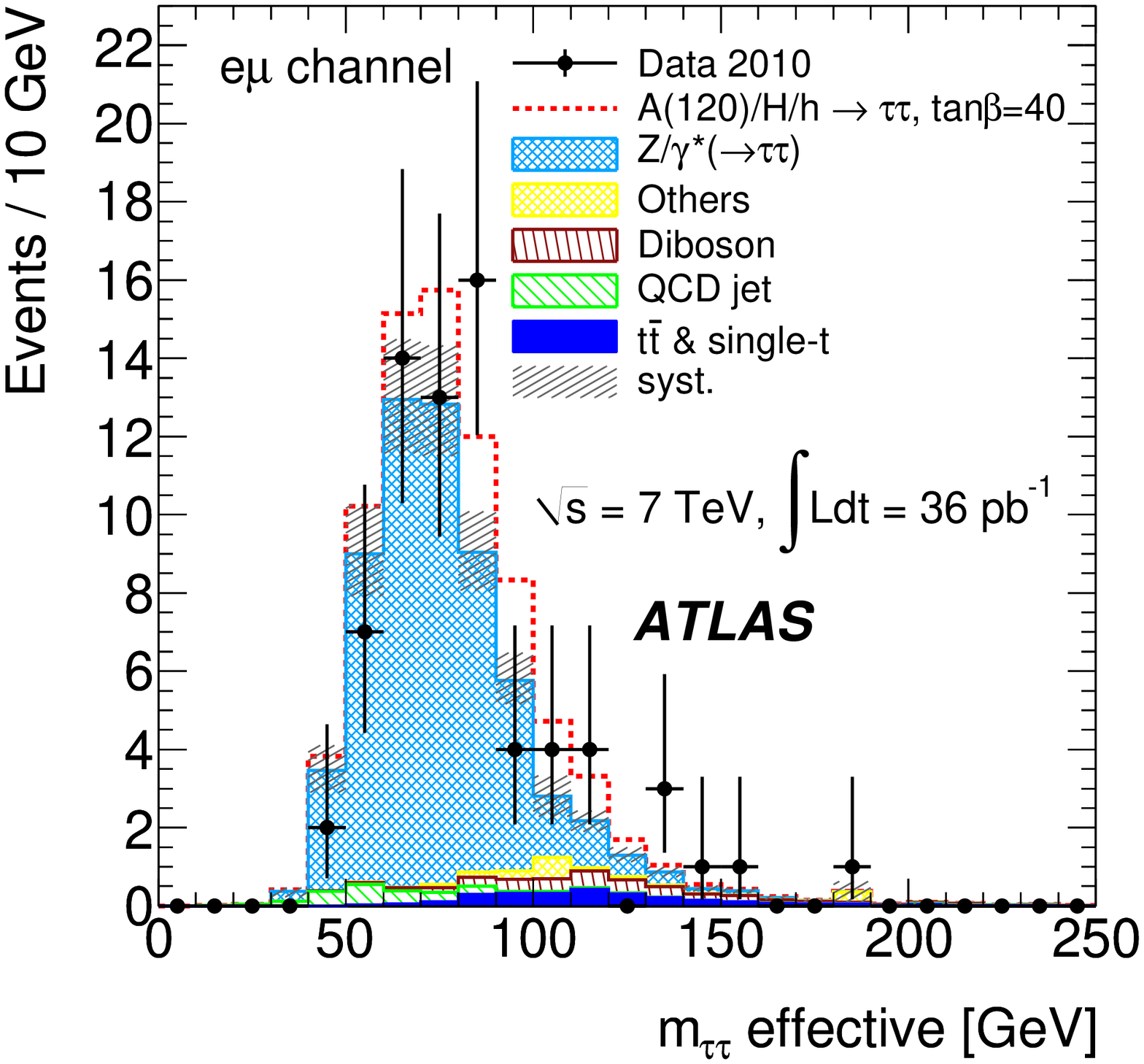}
  \includegraphics[width=0.32\textwidth]{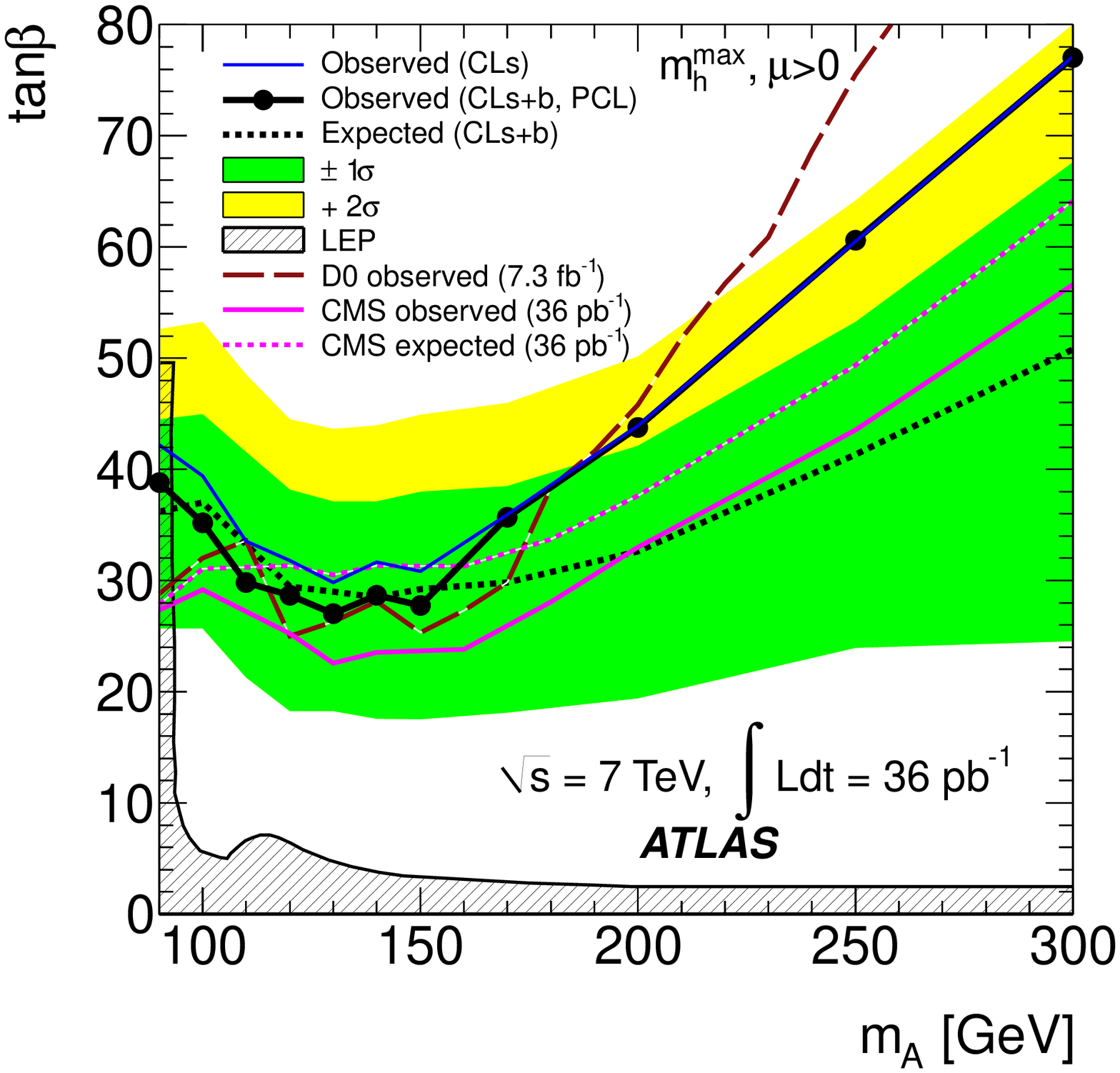}
  \end{center}
\vspace{-0.7cm}
  \caption{Final mass discriminants after the lepton-hadron (left) and lepton-lepton event selections (center). The 
figure to the right shows the resulting 95\% C.L. limits.}
\label{fig:Amass}
\end{figure}

%https://atlas.web.cern.ch/Atlas/GROUPS/PHYSICS/PAPERS/HIGG-2011-02/HIGG-2011-02.pdf
The lepton-hadron event selection requires exactly one electron or muon and one 
oppositely-charged $\tau$ jet, significant $E_T^{\mathrm{miss}}$, and a transverse 
mass of the $\ell-E_T^{\mathrm{miss}}$ system below 30 GeV. 
The final discriminant 
after event selection, the visible mass (invariant mass of electron or muon and the $\tau$ jet), 
is shown in Fig.~\ref{fig:Amass}.
In the lepton-lepton analysis, an isolated electron and an isolated muon with opposite charge 
and an opening angle larger than 2.0 rad are required. Additionally, the scalar 
sum of the transverse momenta of the leptons and of the missing transverse momentum must 
be smaller than 120 GeV.
The final discriminant 
after event selection, the effective mass $m_{\tau\tau}^\mathrm{effective}=\sqrt{(p_e+p_\mu+p_\mathrm{miss})^2}$, 
is shown in Fig.~\ref{fig:Amass}.
The observation is compatible with the data, thus exclusion limits are set (see Fig.~\ref{fig:Amass}), 
extending existing Tevatron limits for $m_A>180$ GeV.

%\begin{figure}[!h!tpb]
%  \begin{center}
%  \includegraphics[width=0.40\textwidth]{figures/neut_limit.eps}
%  \includegraphics[width=0.40\textwidth]{figures/neut_limit_sm.eps}
%  \end{center}
%\vspace{-0.7cm}
%  \caption{A limit}
%\label{fig:Alimit}
%\end{figure}

%\cite{newtautau}

\section{Conclusions}
\noindent
Results from all ATLAS searches for MSSM Higgs bosons with up to 36 fb$^{-1}$ are compatible with the 
SM expectation. Hence exclusion limits have been set in the $H^+ \to cs$ and $h/H/A \to \tau\tau$ 
channels, confirming and extending (for $m_A>180$ GeV) existing Tevatron limits.

\end{document}